\documentclass[aip,amsmath,amssymb,reprint]{revtex4-1}
\usepackage{graphicx}
\usepackage{dcolumn}
\usepackage{bm}

\usepackage[utf8]{inputenc}
\usepackage[T1]{fontenc}
\usepackage{mathptmx}
\usepackage{natbib}
\usepackage{comment}

\draft 

\newcommand*{\Fig}[1]{Fig.~\ref{#1}}
\newcommand*{\GHz}{~\mathrm{GHz}}

\newcommand*{\mm}{~\mathrm{mm}}

\newcommand*{\K}{~\mathrm{K}}

\newcommand*{\mWmK}{~\mathrm{mW}/\mathrm{m}\cdot\mathrm{K}}

\begin{document}

\title{Production Method of Millimeter-Wave Absorber with 3D-Printed Mold} 

\author{S. Adachi}
\email[]{adachi.syunsuke.52e@st.kyoto-u.ac.jp}
\affiliation{Division of Physics and Astronomy, Graduate School of Science, Kyoto University, Kitashirakawa, Sakyo, Kyoto, 606-8502, Japan}
\author{M. Hattori}
\affiliation{Astronomical Institute, Graduate School of Science, Tohoku University, 6-3, Aramaki Aza-Aoba, Aoba-ku, Sendai 980-8578, Japan}
\author{F. Kanno}
\affiliation{Astronomical Institute, Graduate School of Science, Tohoku University, 6-3, Aramaki Aza-Aoba, Aoba-ku, Sendai 980-8578, Japan}
\author{K. Kiuchi}
\affiliation{Department of Physics, Graduate School of Science, The University of Tokyo, 7-3-1 Hongo, Bunkyo-ku, Tokyo 113-0033, Japan}
\author{T. Okada}
\affiliation{Astronomical Institute, Graduate School of Science, Tohoku University, 6-3, Aramaki Aza-Aoba, Aoba-ku, Sendai 980-8578, Japan}
\author{O. Tajima}
\affiliation{Division of Physics and Astronomy, Graduate School of Science, Kyoto University, Kitashirakawa, Sakyo, Kyoto, 606-8502, Japan}

\date{\today}

\begin{abstract}
We established a production method of a millimeter-wave absorber by using a 3D-printed mold.~\ %
The mold has a periodic pyramid shape,~\
and an absorptive material is filled into the mold.~\
This shape reduces the surface reflection.~\
The 3D-printed mold is made from a transparent material in the millimeter-wave range.~\
Therefore, unmolding is not necessary.~\
A significant benefit of this production method is easy prototyping with various shapes and various absorptive materials.~\
We produced a test model~\
and used a two-component epoxy encapsulant as the absorptive material.~\
The test model achieved a low reflectance:~$\sim1\%$ at $100\GHz$.~\
The absorber is sometimes maintained at a low temperature condition for cases in which superconducting detectors are used.~\
Therefore, cryogenic performance is required in terms of~\
a mechanical strength for the thermal cycles, an adhesive strength, and a sufficient thermal conductivity.~\
We confirmed the test-model strength by immersing the model into a liquid-nitrogen bath.~\
\end{abstract}

\pacs{}

\maketitle 

\begin{figure*}
  \centering

  \includegraphics[width=0.30\textwidth]{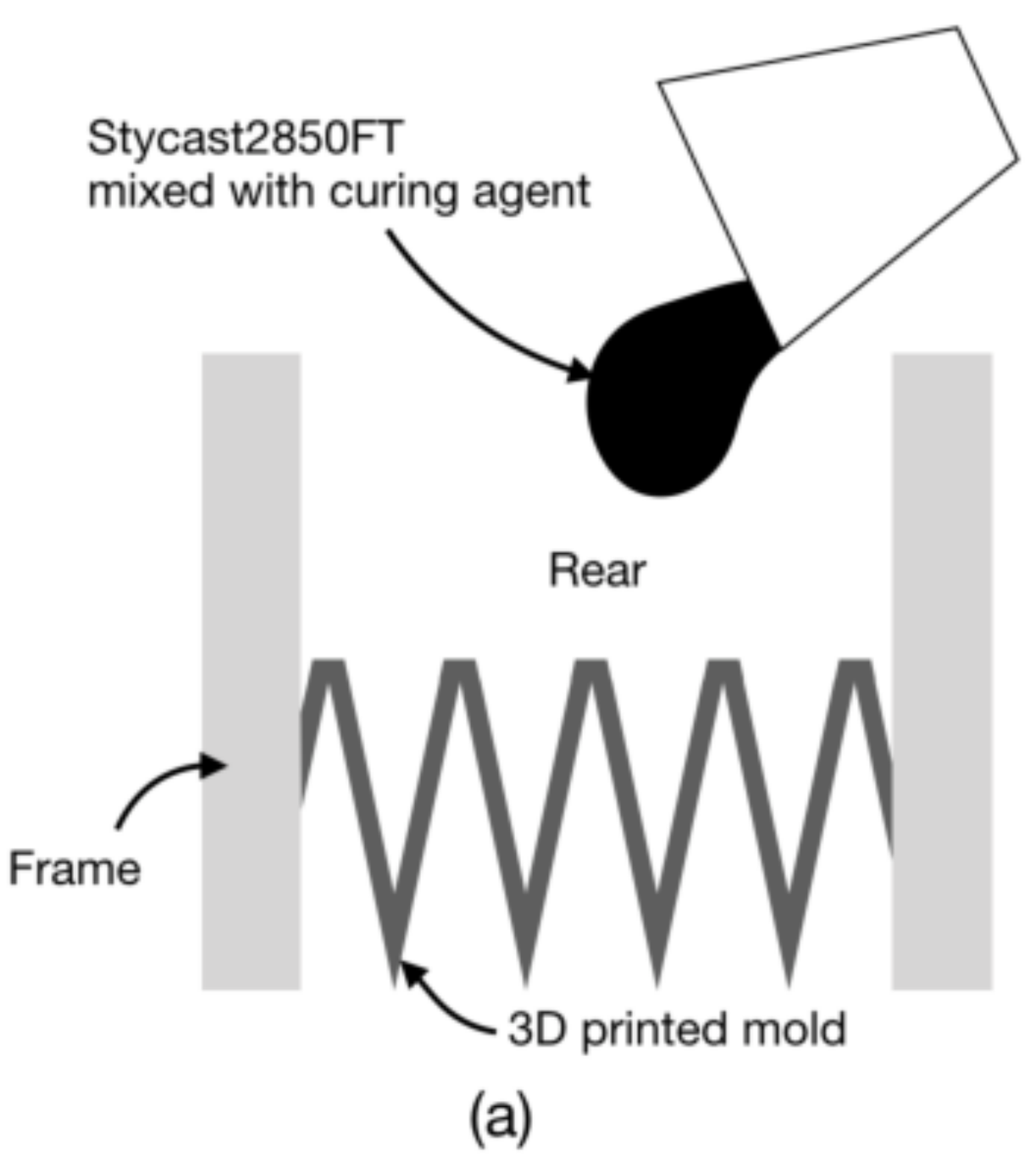}
  \includegraphics[width=0.30\textwidth]{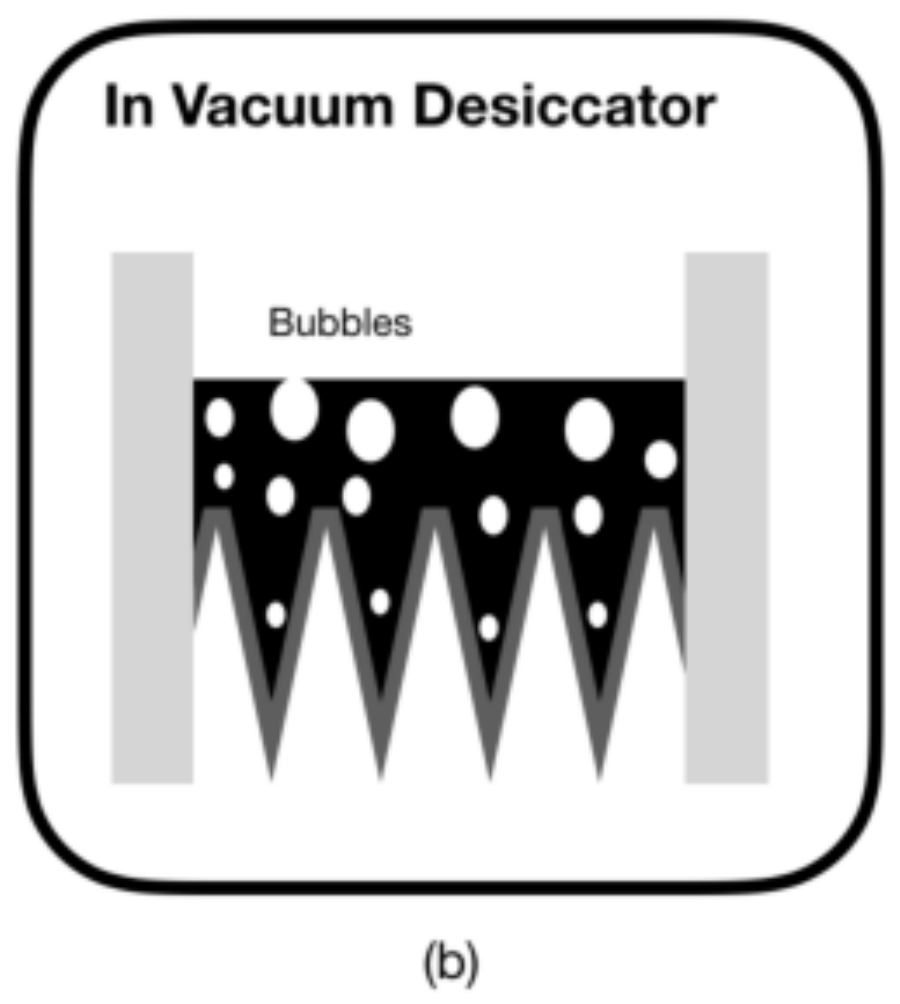}
  \includegraphics[width=0.24\textwidth]{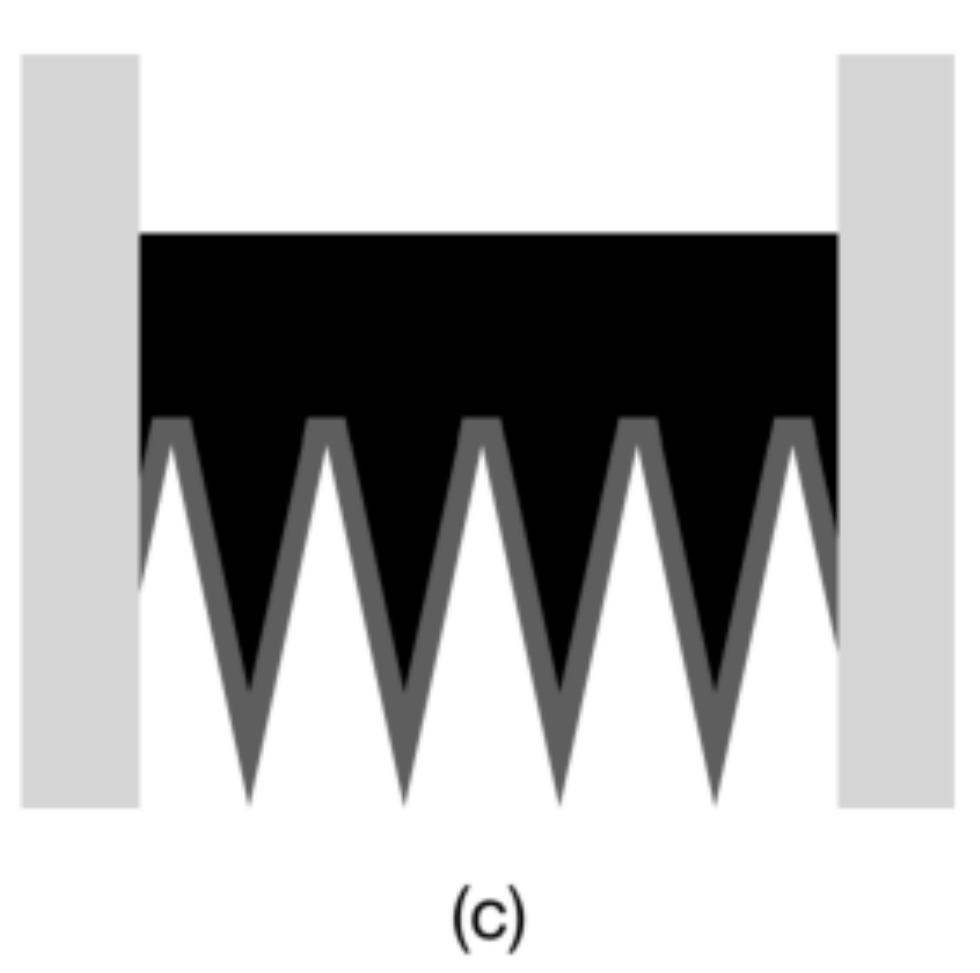}

  \caption{
    Production procedure for the absorber:~\
    (a) pouring Stycast2850FT mixed with curing agent into the 3D-printed mold,
    (b) degassing in a vacuum desiccator,
    and (c) curing at room temperature.
    We fill the absorptive material from the rear of the 3D-printed mold.~\
    \label{procedure}}%
\end{figure*}

Superconductive detectors are used extensively~\
for high-precision measurements of millimeter-waves.~\
In such a measurement system,~\
the reduction of stray light, i.e., undesirable rays from unintended optical paths, is essential.~\
The light path is composed of multiple reflections on the inner wall of the optics system.~\
Therefore, a millimeter-wave absorber installation on the wall is a promising strategy to reduce the stray light\cite{PB2,ACT,BICEP2}.~\
Because the optics system is commonly cooled to suppress thermal noise,~\
the absorber is required to have a good cryogenic performance:~\
a mechanical strength for the thermal cycles, an adhesive strength, and a sufficient thermal conductivity.~\

A mixture of epoxy and carbon or stainless steel is used commonly as an absorber for low-temperature applications.~\
Stycast2850FT~(Henkel Corporation), which is a two-component epoxy encapsulant, is used commonly\cite{arcade,ACT}.~\
Its index of reflection\cite{Stycast} is $n\sim2.3$.~\
Its surface reflectivity is $\sim15\%$.~\
Further reduction of the surface reflectivity can be achieved by forming surface structures,~\
such as pyramids and needles\cite{yokomori,THzAbsorber}.~\
Molding is the most popular production method for making the surface structure\cite{moldPyramid}.~\
However, a long lead-time is required to make a mold.~\
Moreover, it is difficult to unmold the absorptive material because of its strong adhesion.~\


\begin{figure}
\includegraphics[width=0.35\textwidth]{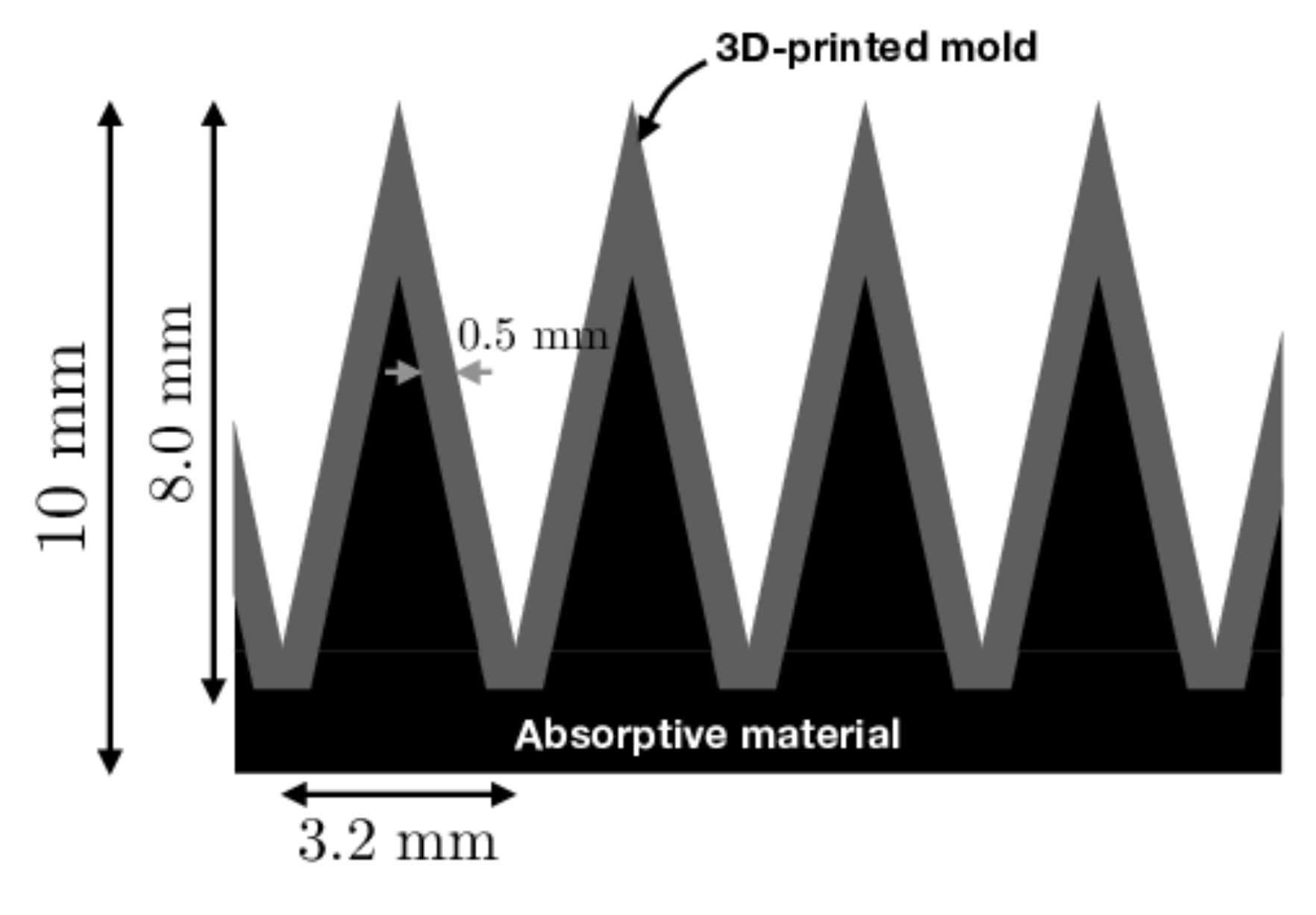}%
\caption{
  Cross-sectional schematic of the test model.~\
  \label{3Dshape}}%
\end{figure}

To produce a pyramidal-shaped absorber easily,~\
we established a new production method for the absorber with a 3D-printed mold.~\
This method does not require an unmolding step if the 3D-printed mold is made from a transparent material in the millimeter-wave range.~\
The surface structure of the periodic quadrangular pyramid is shaped with the 3D-printed mold.~\
As illustrated in \Fig{procedure},~\
we fill the absorptive material from the rear of the mold.~\
In a previous study\cite{3DprinterAbsorber},~\
the absorber is made by the 3D printer.~\
In this method, the choice of the absorptive material is limited.
Various materials can be used in our method. 
It is easy to prototype the absorber with various shapes and various materials.~\
This benefit also allows us to optimize the cryogenic performance, such as the thermal conductivity.~\

\begin{figure}
  \includegraphics[width=0.18\textwidth]{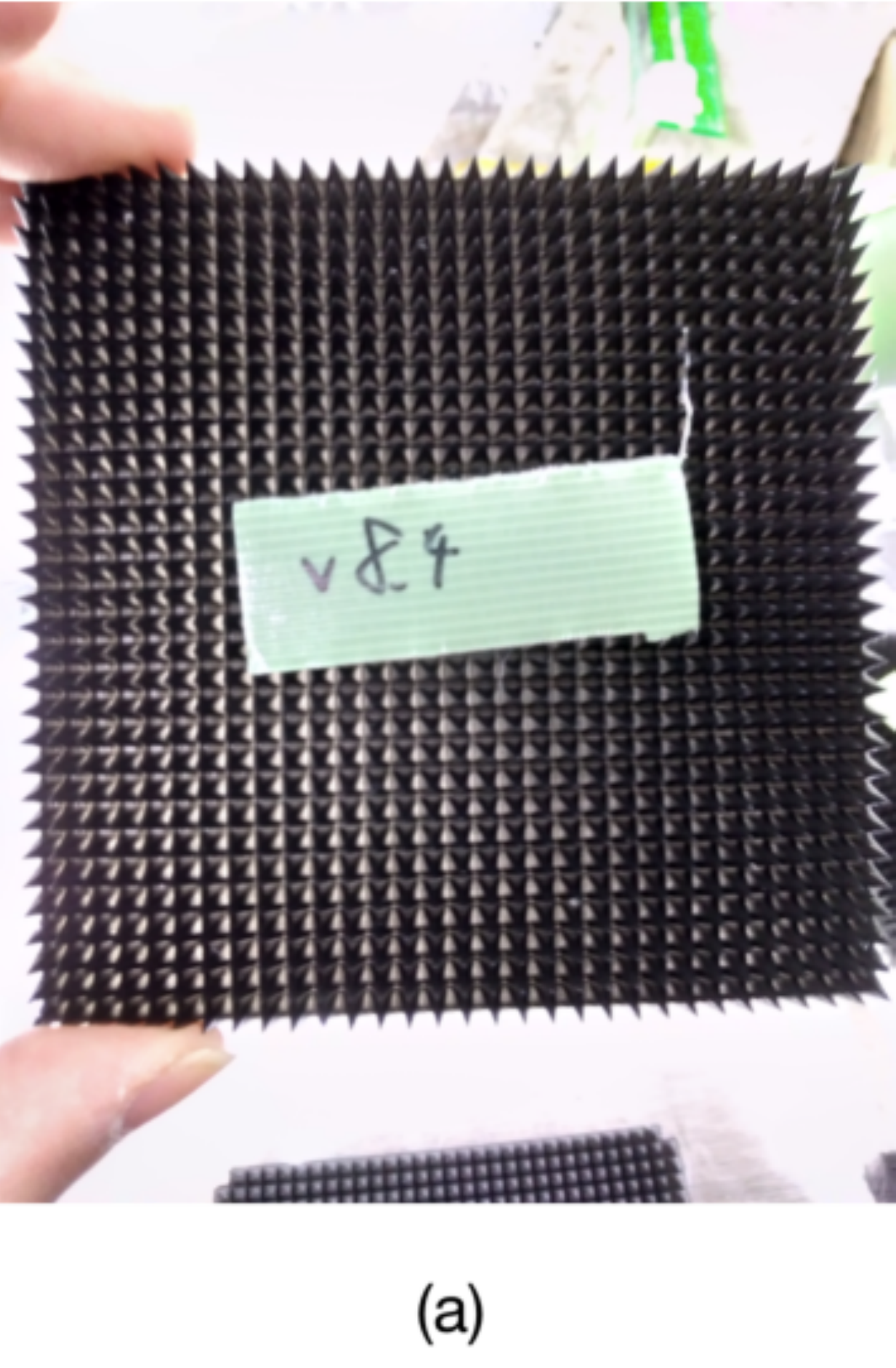}
  \vspace{1ex}
  \includegraphics[width=0.18\textwidth]{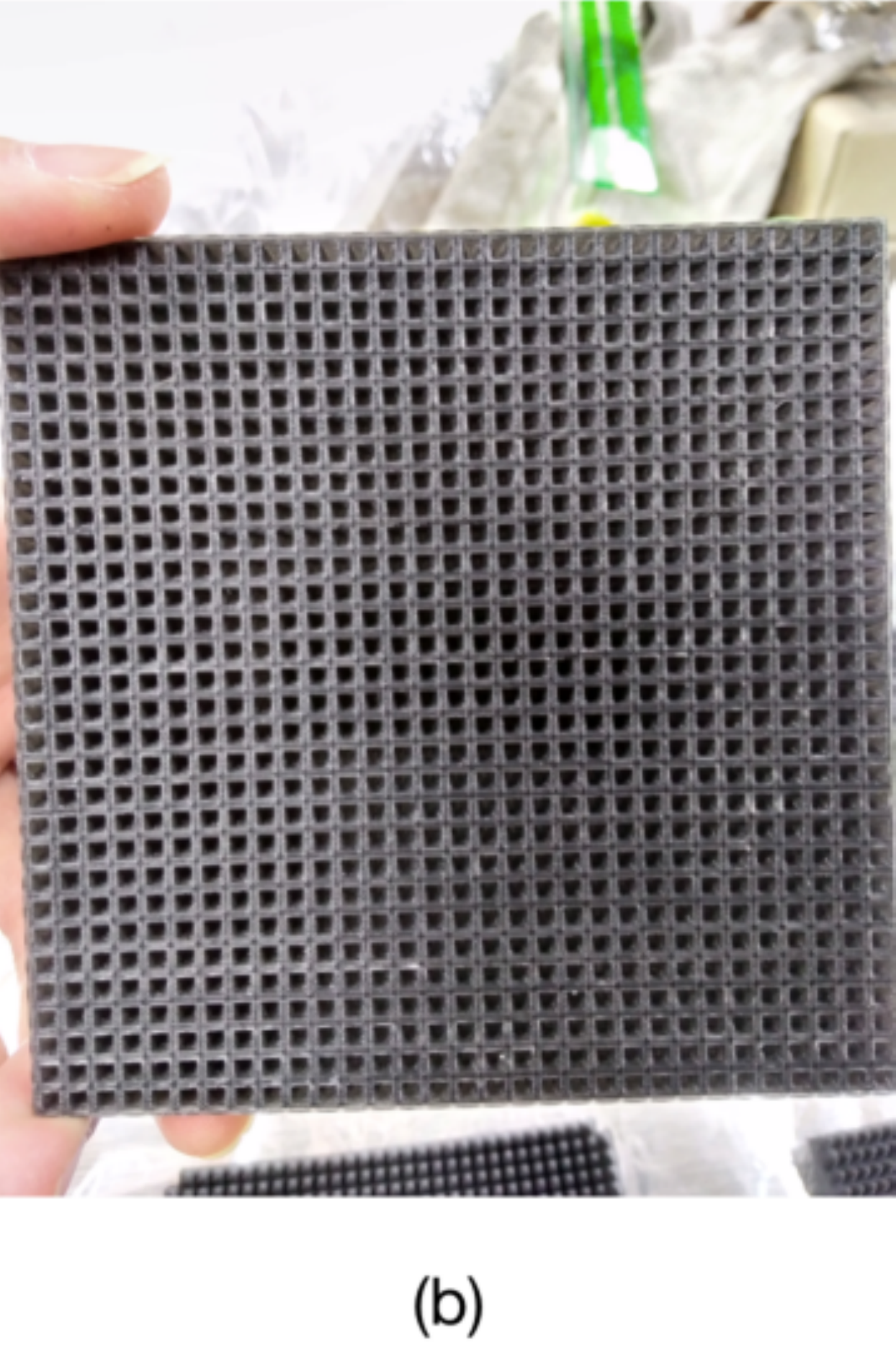}
  \caption{
    Photographs of the test mold with a $100\mm \times 100\mm$ area.~\
    (a) is the front view, and (b) is the rear view.~\
    \label{3Dpicture}}%
\end{figure}

To demonstrate the usefulness of our production method, we produced a test model.~\
For the 3D-printing, we used the PolyJet method~(Stratasys Japan Ltd.),~\
which was an analogous method that was similar to ink-jet paper-printing.~\
The mold material was VeroBlack.~\
The mold geometry can be controlled with a printing fineness of $0.03\mm$,~\
which is several times better than the conventional 3D-printing method\cite{Vero,FDM}.~\
As illustrated in \Fig{3Dshape}, we made quadrangular pyramids with a height and width~\
of $8.0\mm$ and $3.2\mm$, respectively.~\
The mold thickness was $0.5\mm$.~\
We used the same aspect ratio as in a previous study~\
for the cosmic microwave background project\cite{arcade};~\
the height over the pitch was $2.3$.~\
Photographs of the test mold are shown in \Fig{3Dpicture}.~\

\begin{figure}
  \includegraphics[width=0.35\textwidth]{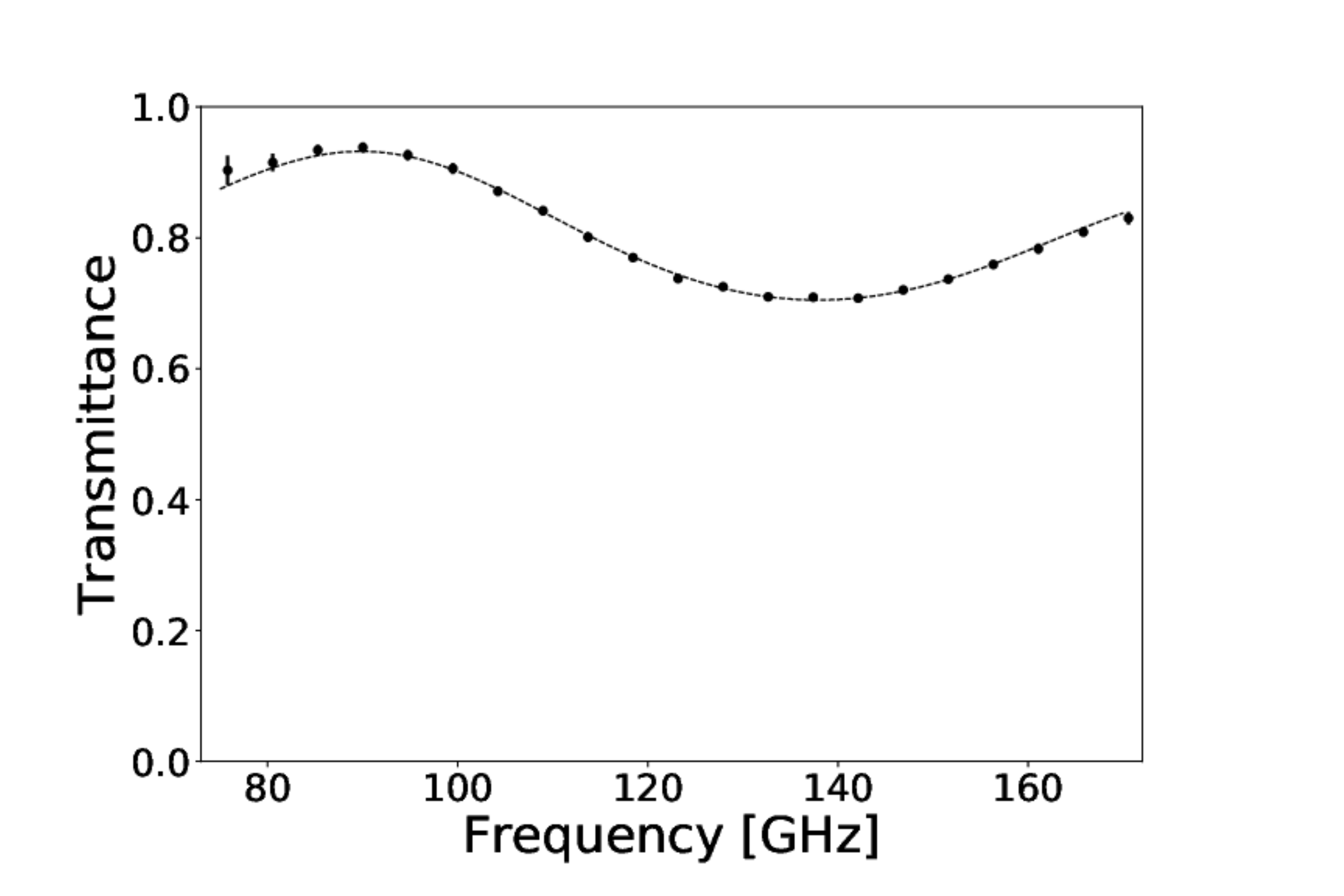}
  \caption{
    Measured transmittance of VeroBlack flat sheet~($1\mm$-thickness).~\
    A dashed line is a fit result to the data with floating optical parameters~(see text for details).~\
    \label{VeroBlack}}%
\end{figure}

We measured the transmittance of the VeroBlack flat sheet ($1\mm$-thickness) from $75\GHz$ to $170\GHz$.~\
We used a Martin--Puplett-type Fourier transform spectrometer~(FTS)\cite{hattoriFTS}$^\mathrm{,}$\
\footnote{This FTS works according to the principle of an auto-correlation mode of a multi-Fourier-transform interferometer.}\
with a semiconducting bolometer\cite{HattoriBolometer}.~\ 
Figure~\ref{VeroBlack} shows the measured transmittance at normal incidence.~\
We repeated the measurements 20 times and assigned errors for each point with their standard deviations.~\
The transmittance was $\geq70\%$ in the measured frequency range.~\
By fitting the data, we extracted its optical parameters:~\
the index of refraction, loss tangent, and the plate thickness,~\
which were $1.70\pm0.05$, $0.020\pm0.005$, and $0.97\pm0.03\mm$, respectively.~\
These results indicate that the absorption loss of the mold is sufficiently low~\
and the mold can be used as a transparent material.

\begin{figure}
  \includegraphics[width=0.4\textwidth]{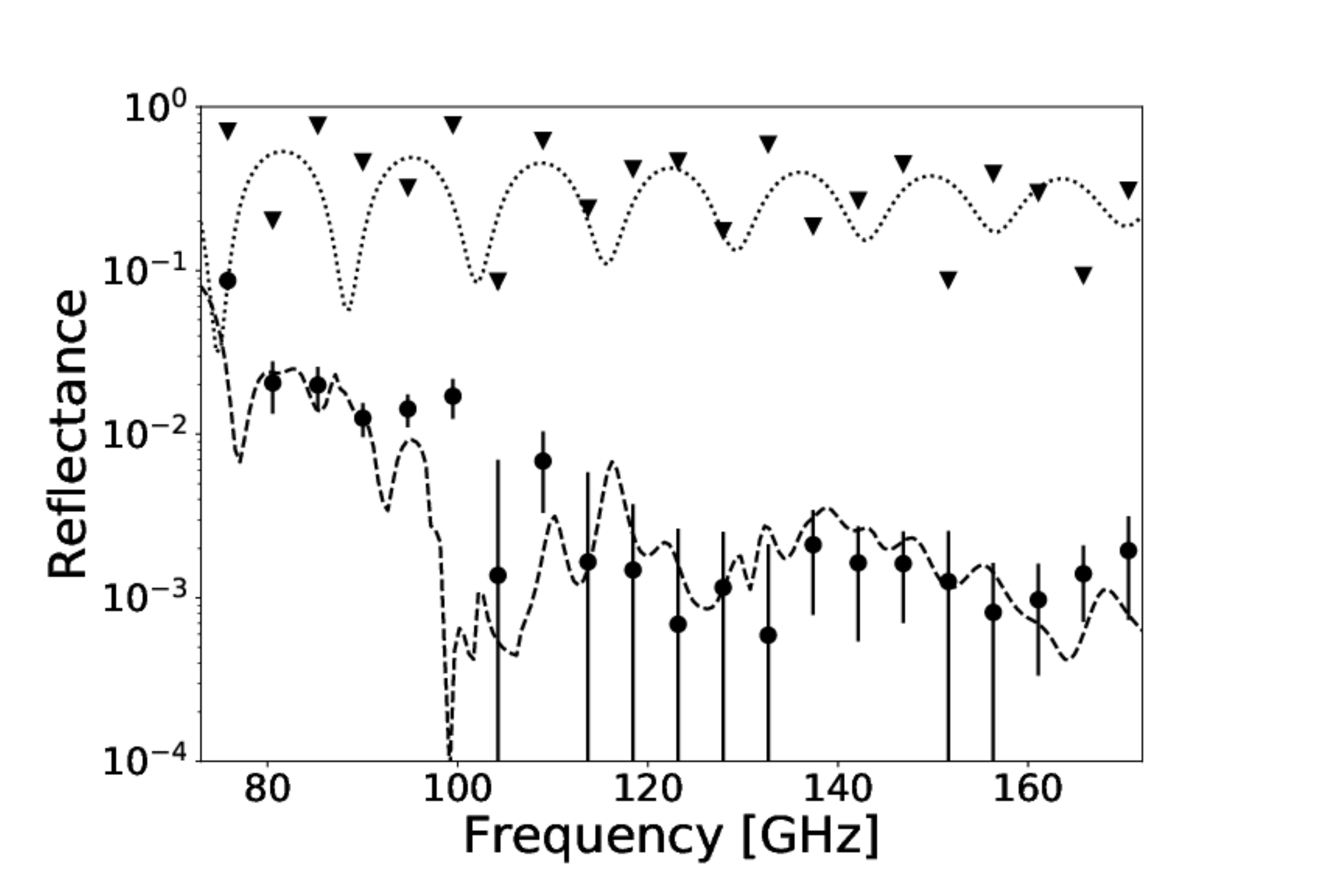}
\caption{
  Measured reflectance of test model~(circle) and flat Stycast2850FT~(triangle) with incident angle of $45^{\circ}$.
  Dashed and dotted lines are simulation results for each case.
  \label{result}}%
\end{figure}

For the absorptive material, we used Stycast2850FT.~\
By using the FTS described above, we measured the reflectance of the test model and flat Stycast2850FT ($5\mm$-thickness).~\
The samples were set before an aluminum plate.~\
The incident angle of the light to the samples was $45^{\circ}$.~\
We measured the reflected power at $45^{\circ}$.~\
Figure~\ref{result} shows the measured reflectance as a function of frequency.~\
The periodic pyramid structure reduced the reflectance significantly~($\sim1\%$ at $100\GHz$)~\
compared with the case of the flat surface~($\approx50\%$ at $100\GHz$).~\
Based on catalog numbers for the index of refraction~($2.32$) and the loss tangent~($0.051$) of Stycast2850FT\cite{Stycast}, we performed a simulation~\
by using the ANSYS-HFSS~[ANSYS Inc.].~\
We confirmed the reason for the improvement from the consistency between the measured and simulation results.~\

\begin{figure}
  \includegraphics[width=0.25\textwidth]{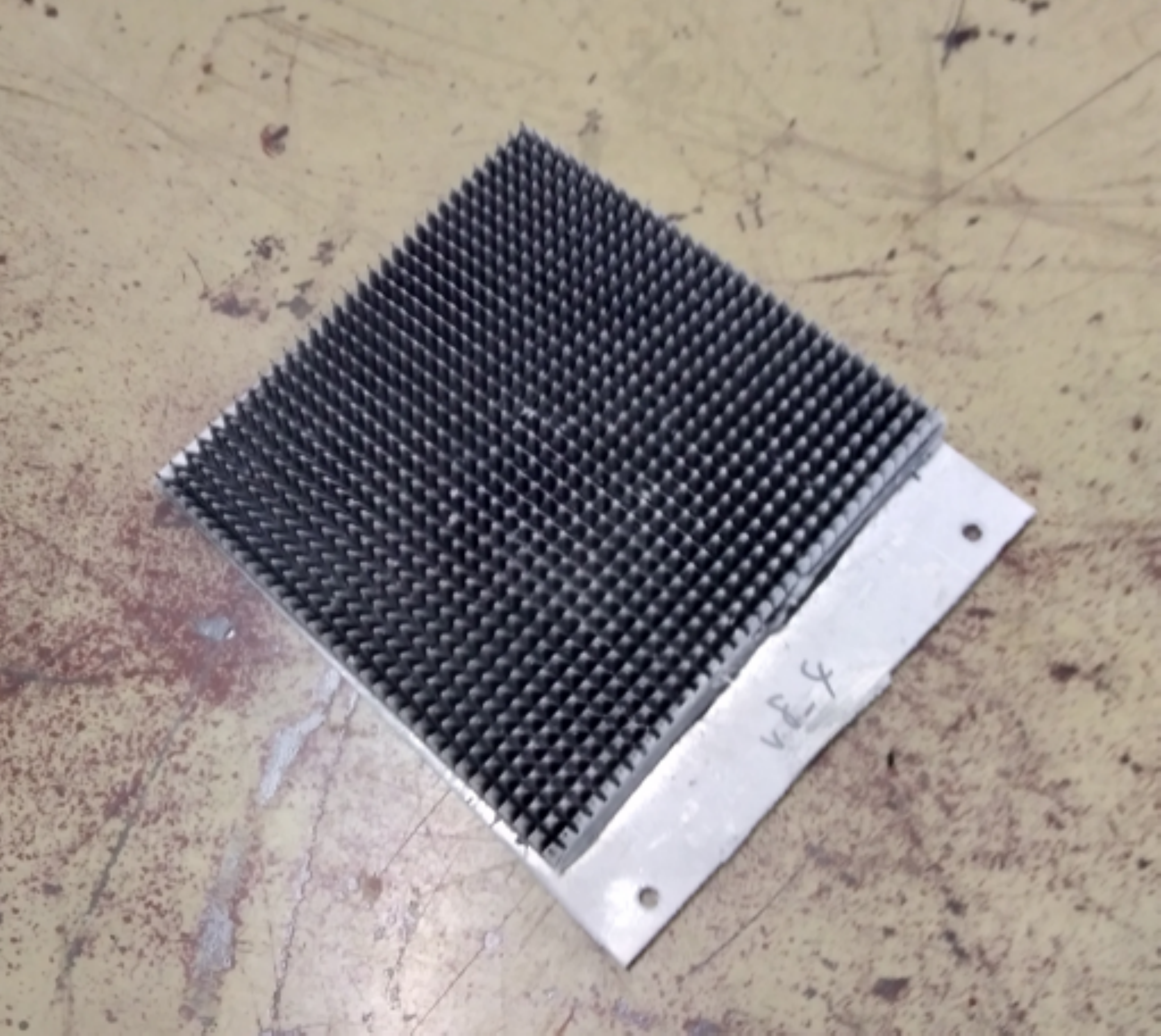}%
  \caption{
    Photograph after immersion into liquid nitrogen three times.~\
    No cracks were found on its surface, which confirmed the strong adhesion.~\
    \label{LN2}}%
\end{figure}

The test model is aimed primarily towards cryogenic application.~\
The coefficient of thermal expansion~(CTE) of the Stycast2850FT matches that of the aluminum.~\
Stycast2850FT has a good adhesive performance at a low-temperature condition,~\
and it is commonly used as an adhesive in cryostats\cite{adhesive,StycastCommon}.~\
Stycast2850FT also has a sufficient thermal conductivity of $\sim50\mWmK$ at $4\K$~\cite{StycastTC}.~\
The CTE of the 3D-printed resin~($\sim100\mathrm{ppm}/\mathrm{K}$)\cite{resinCTE} is larger than that of the Stycast2850FT~($\sim40\mathrm{ppm}/\mathrm{K}$)\cite{Stycast}.~\
Because the CTE difference may cause a mechanical stress, we performed a stress test.~\
We immersed the absorber that adhered to the aluminum plate into liquid nitrogen~($77\K$) three times.~\
Figure~\ref{LN2} shows a photogragh of the absorber after the immersion test.~\
We do not find any cracks on its surface, and we confirmed the good adhesive performance.~\
The immersion test generated a significantly higher thermal stress than in the actual applications.~\
Ordinarily, cooling takes a long time~($\gg 1$~hour).~\

An important benefit of the 3D-printed mold absorber is~\
easy prototyping for various applications,~\
which is difficult for previous molding production.~\
It is easy to change the shape and the absorptive material.~\
A material exists with a larger loss tangent than the Stycast2850FT\cite{CarbonEpoxy}.~\
Further improvements are expected.~\

In summary, we established a novel production method for the millimeter-wave absorber~\
by filling absorptive material into the 3D-printed mold.~\
The mold is thin and transparent in the radiofrequency range.~\
The approach provides an easy method to make a pyramidal-textured absorber.~\
This structure reduces the reflectance significantly, which is the most important parameter for the absorber.~\
Based on this method, we produced a test model with Stycast2850FT as the absorptive material.
We confirmed its low reflectance~($\sim1\%$ at $100\GHz$).~\
We also confirmed its cryogenic performance:~\
a mechanical strength for the thermal cycles, an adhesive strength, and a sufficient thermal conductivity.~\
Flexibilities for changing the geometry and the absorptive material are~\
convenient to develop absorbers for new applications.~\

\begin{acknowledgments}
We would like to thank the Stratasys Japan Ltd. for fabrication of the 3D-printed mold.~\
We thank Shugo Oguri and Taketo Nagasaki for assistance with the optics simulations and Minoru Hirose for technical assistance with the absorber production.~\
We also thank Akito Kusaka and Frederick T. Matsuda for the useful discussion.~\
This work was supported by JSPS KAKENHI Grant Number JP18J01039, JP17H06134, and SPIRITS which is the internal grant of Kyoto University.

\end{acknowledgments}

\bibliography{submit_v3}

\end{document}